\documentclass[fleqn,a4paper]{vch-book}
\usepackage{amsmath}
\usepackage{amssymb}
\usepackage{cite}
\usepackage{makeidx}
    \makeindex
\usepackage{times}
\usepackage{tabularx}
\usepackage{booktabs}
\usepackage[dvips]{epsfig}
\DeclareGraphicsExtensions{.eps,.ps}
\newcommand{\void}[1]{}
\graphicspath{{./img/}}
\chardef\bslash=`\\ % p. 424, TeXbook
\newcommand{\bibtex}{\ifx\is@itshape\f@shape{\fontshape{scit}\selectfont
Bib}\else\textsc{Bib}\fi\kern-.1em\TeX}

\begin{document}

%\tableofcontents

\chapter{Decoherence in resonantly driven bistable systems}
\authorafterheading{Sigmund Kohler and Peter H\"anggi}
\affil{ Institut f\"ur Physik\\
	Universit\"at Augsburg\\
	86135 Augsburg, Germany}

%----------------------------------------------------------------------
\section{Introduction}

A main obstacle for the experimental realization of a quantum computer is the
unavoidable coupling of the qubits to external degrees of freedom and the
decoherence caused in that way.  A possible solution of this problem are error
correcting codes.  These, however, require redundant coding and, thus, a
considerably higher algorithmic effort.

Yet another route to minimize decoherence is provided by the use of
time-dependent control fields.  Such external fields influence the coherent and
the dissipative behavior of a quantum system and can extend coherence times
significantly.  One example is the stabilization of a coherent superposition in
a bistable potential by coupling the system to an external dipole field
\cite{Dittrich1993a,Grifoni1998a}.  The fact that a driving field reduces the
effective level splitting and therefore decelerates the coherent dynamics as
well as the dissipative time evolution is here of cruical influence.  A qubit
is usually represented by two distinguished levels of a more complex quantum
system and, thus, a driving field may also excite the system to levels outside
the doublet that forms the qubit, i.e., cause so-called leakage.  While a small
leakage itself may be tolerable for the coherent dynamics, its influence on the
quantum coherence of the system may be even more drastic.  We demonstrate in
this article that in a drivien qubit resonances with higher states, which are
often ignored, may in fact enhance decoherence substantially.

A related phemomenon has been found in the context of dissipative chaotic
tunneling near singlet-doublet crossings where the influence of so-called
chaotic levels yields an enhanced loss of coherence
\cite{Kohler1998a,Hanggi1999a}.

%----------------------------------------------------------------------
\section{The model and its symmetries}

\begin{figure}[tb]
\includegraphics[width=.5\columnwidth]{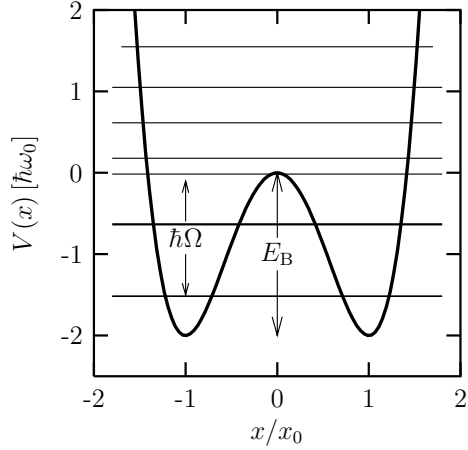}
\hfill\parbox[b]{.45\columnwidth}{
\caption{\label{ddw:fig:potential}
Sketch of the double well potential in Eq.~(\ref{hamiltonian})
for $D=E_\mathrm{B}/\hbar\omega_0=2$.
The horizontal lines mark the eigenenergies in the absence
of the driving; the levels below the barrier come in doublets.}
\vspace{5ex}}
\end{figure}

We consider as a working model the quartic double well with a spatially
homogeneous driving force,
harmonic in time. It is defined by the Hamiltonian
\begin{equation}
H(t) = \frac{p^2}{2m} - \frac{1}{4}m\omega_0^2x^2
       + \frac{m^2\omega_0^4}{64E_\mathrm{B}}x^4 
       + Sx\cos(\Omega t) . \label{hamiltonian}
\end{equation}
The potential term of the static bistable Hamiltonian, $H_\mathrm{DW}$,
possesses two minima at $x=\pm x_0$, $x_0=(8E_\mathrm{B}/m\omega_0^2)^{1/2}$,
separated by a barrier of height $E_\mathrm{B}$ (cf.\
Fig.~\ref{ddw:fig:potential}).  The parameter $\omega_0$ denotes the (angular)
frequency of small oscillations near the bottom of each well.  Thus, the energy
spectrum consists of approximately $D=E_\mathrm{B}/\hbar\omega_0$ doublets
below the barrier and singlets which lie above.  As a dimensionless measure for
the driving strength we use $F=S(8m\omega_0^2E_\mathrm{B})^{-1/2}$.
%----------
%----------

The Hamiltonian (\ref{hamiltonian}) is $T$-periodic, with $T = 2\pi/\Omega$. As
a consequence of this discrete time-translational invariance of $H(x,p;t)$, the
relevant generator of the quantum dynamics is the one-period propagator
\cite{Grifoni1998a,Shirley1965a,Sambe1973a,Manakov1986a,QTAD}
\begin{equation}
U(T,0) = {\cal T}\,\exp\left( -\frac{\mathrm{i}}{\hbar}
\int_0^T \mathrm{d} t \, H_\mathrm{DW}(t) \right),
\end{equation}
where ${\cal T}$ denotes time ordering. According to the Floquet theorem, the
Floquet states of the system are the eigenstates of $U(T,0)$.
They can be written in the form
\begin{equation}
|\psi_{\alpha}(t)\rangle = \mathrm{e}^{-\mathrm{i} \epsilon_{\alpha}t/\hbar}
|\phi_{\alpha}(t)\rangle, \quad
\label{eq:floquetstat}
\end{equation}
with
\begin{eqnarray*}
|\phi_{\alpha}(t + T)\rangle = |\phi_{\alpha}(t)\rangle.
\end{eqnarray*}
Expanded in these Floquet states, the propagator of the driven system reads
\begin{equation}
U(t,t')=\sum_\alpha \mathrm{e}^{-\mathrm{i}\epsilon_\alpha (t-t')/\hbar}
|\phi_\alpha(t)\rangle\langle\phi_\alpha(t')|.
\label{eq:floquetprop}
\end{equation}
The associated eigenphases $\epsilon_{\alpha}$, referred to as quasienergies,
come in classes, $\epsilon_{\alpha,k}=\epsilon_{\alpha}+k\hbar\Omega$,
$k=0,\pm 1,\pm 2,\ldots$. This is suggested by a Fourier expansion of the
$|\phi_{\alpha}(t)\rangle$,
\begin{eqnarray}
|\phi_{\alpha}(t)\rangle
&=& \sum_k |\phi_{\alpha,k}\rangle\,\mathrm{e}^{-\mathrm{i}k\Omega t}, \nonumber \\
|\phi_{\alpha,k}\rangle
&=& \frac{1}{T} \int_0^T \mathrm{d} t \, |\phi_{\alpha}(t)\rangle\,
\mathrm{e}^{\mathrm{i} k\Omega t}.
\label{floquet:fourier}
\end{eqnarray}
The index $k$ counts the number of quanta in the driving field. Otherwise, the
members of a class $\alpha$ are physically equivalent. Therefore, the
quasienergy spectrum can be reduced to a single ``Brillouin zone'',
$-\hbar\Omega/2 \leq \epsilon < \hbar\Omega/2$.
 
Since the quasienergies have the character of phases, they can be ordered only
locally, not globally. A quantity that is defined on the full real axis and
therefore does allow for a complete ordering, is the mean energy
\cite{Grifoni1998a,QTAD}
\begin{equation}
E_{\alpha}
= \frac{1}{T} \int_0^T \mathrm{d} t \,
\langle\psi_{\alpha}(t)|\,H_\mathrm{DW}(t)\,|\psi_{\alpha}(t)\rangle
\label{eq:meanen}
\end{equation}
It is related to the corresponding quasienergy by
\begin{equation}
E_{\alpha} = \epsilon_{\alpha} + \frac{1}{T}\int_0^T \mathrm{d}t\,
\langle\phi_{\alpha}(t)|\,\mathrm{i}\hbar\frac{\partial}{\partial
t}\,|\phi_{\alpha}(t)\rangle  .
\end{equation}
Without the driving, $E_{\alpha} = \epsilon_{\alpha}$, as it should be.  By
inserting the Fourier expansion (\ref{floquet:fourier}), the mean energy takes
the form
\begin{equation}
E_\alpha = \sum_k(\epsilon_\alpha+k\hbar\Omega)\,
\langle \phi_{\alpha,k}|\phi_{\alpha,k}\rangle.
\label{eq:meanen:fourier}
\end{equation}
This form reveals that the $k$th Floquet channel yields a contribution
$\epsilon_\alpha + k\hbar\Omega$ to the mean energy, weighted by the Fourier
coefficient $\langle \phi_{\alpha,k}|\phi_{\alpha,k}\rangle$.  For the
different methods to obtain the Floquet states, we refer the reader to the
reviews \cite{QTAD,Grifoni1998a}, and the references therein.

%-------------------------------------
%\subsubsection{Generalized parity.}
%
The invariance of the static Hamiltonian under parity ${\sf P}:(x,p,t)\to
(-x,-p,t)$ is violated by the dipole driving force. With the above choice of
the driving, however, a more general, dynamical symmetry remains.  It is
defined by the operation \cite{Grifoni1998a,QTAD}
\begin{equation}
{\sf P}_T:(x,p,t)\to(-x,-p,t+T/2)
\label{parity}
\end{equation}
and represents a generalized parity acting in the extended phase space spanned
by $x$, $p$, and phase, i.e., time $t\mathop{\mathrm{mod}}T$.  While such a
discrete symmetry is of minor importance in classical physics, its influence on
the quantum mechanical quasispectrum $\{\epsilon_\alpha(S,\Omega)\}$ is
profound: It devides the Hilbert space in an even and an odd sector, thus
allowing for a classification of the Floquet states as even or odd.
Quasienergies from different symmetry classes may intersect, while
quasienergies with the same symmetry typically form avoided crossings. The fact
that ${\sf P}_T$ acts in the phase space extended by time
$t\mathop{\mathrm{mod}}T$, results in a particularity: If, e.g.,
$|\phi(t)\rangle$ is an even Floquet state, then $\exp(\mathrm{i}\Omega
t)|\phi(t)\rangle$ is odd, and vice versa.  Thus, two equivalent Floquet states
from neighboring Brillouin zones have opposite generalized parity. This means
that a classification of the corresponding solutions of the Schr\"odinger
equation, $|\psi(t)\rangle=\exp(-\mathrm{i}\epsilon t/\hbar)|\phi(t) \rangle$,
as even or odd is meaningful only with respect to a given Brillouin zone.

%----------------------------------------------------------------------
\section{Coherent tunneling}

With the driving switched off, $S=0$, the classical phase space generated by
$H_\mathrm{DW}$ exhibits the constituting features of a bistable Hamiltonian
system: A separatrix at $E = 0$ forms the border between two sets of
trajectories: One set, with $E < 0$, comes in symmetry-related pairs, each
partner of which oscillates in either one of the two potential minima. The
other set consists of unpaired, spatially symmetric trajectories, with $E > 0$,
which encircle both wells.
 
Torus quantization of the integrable undriven double well implies a simple
qualitative picture of its eigenstates: The unpaired tori correspond to
singlets with positive energy, whereas the symmetry-related pairs below the top
of the barrier correspond to degenerate pairs of eigenstates.  Due to the
almost harmonic shape of the potential near its minima, neighboring pairs are
separated in energy approximately by $\hbar\omega_0$. Exact quantization,
however, predicts that the partners of these pairs have small but finite
overlap. Therefore, the true eigenstates come in doublets, each of which
consists of an even and an odd state, $|\Phi_n^+\rangle$ and
$|\Phi_n^-\rangle$, respectively. The energies of the $n$th doublet are
separated by a finite tunnel splitting $\Delta_n$. We can always choose the
global relative phase such that the superpositions
\begin{equation}
|\Phi_n^\mathrm{R,L}\rangle = \frac{1}{\sqrt{2}}\left(
|\Phi_n^+\rangle \pm |\Phi_n^-\rangle \right)
\end{equation}
are localized in the right and the left well, respectively.  As time evolves,
the states $|\Phi_n^+\rangle$, $|\Phi_n^-\rangle$ acquire a relative phase
$\exp(-\mathrm{i}\Delta_n t/\hbar)$ and $|\Phi_n^\mathrm{R}\rangle$,
$|\Phi_n^\mathrm{ L}\rangle$ are transformed into one another after a time
$\pi\hbar/\Delta_n$.  Thus, the particle tunnels forth and back between the
wells with a frequency $\Delta_n/\hbar$. This introduces an additional, purely
quantum-mechanical frequency scale, the tunneling rate $\Delta_0/\hbar$ of a
particle residing in the ground-state doublet.  Typically, tunneling rates are
extremely small compared to the frequencies of the classical dynamics.
 
The driving in the Hamiltonian (\ref{hamiltonian}), even if its influence on
the classical phase space is minor, can entail significant consequences for
tunneling: It may enlarge the tunnel rate by orders of magnitude or even
suppress tunneling altogether. For adiabatically slow driving, i.e.\
$\Omega\ll\Delta_0/\hbar$, tunneling is governed by the instantaneous tunnel
splitting, which is always larger than its unperturbed value $\Delta_0$ and
results in an enhancement of the tunneling rate \cite{Grossmann1991a}.  If the
driving is faster, the opposite holds true: The relevant time scale is now given
by the inverse of the quasienergy splitting of the ground-state doublet
$\hbar/|\epsilon_1-\epsilon_0|$.  It has been found \cite{Grossmann1991a,
Grossmann1991b, Grossmann1992a} that in this case, for finite driving
amplitudes, $|\epsilon_1-\epsilon_0|<\Delta_0$.  Thus tunneling is always
decelerated.  When the quasienergies of the ground-state doublet (which are of
different generalized parity) intersect as a function of $F$, the splitting
vanishes and tunneling can be brought to a complete standstill by the purely
coherent influence of the driving --- not only stroboscopically, but also in
continuous time \cite{Grossmann1991a, Grossmann1991b, Grossmann1992a}.
 
So far, we have considered only driving frequencies much smaller than the
frequency scale $\omega_0$ of the relevant classical resonances.  In this
regime, coherent tunneling is well described within a two-state approximation
\cite{Grossmann1992a}.
Near an avoided crossing, level separations may deviate vastly, in both
directions, from the typical tunnel splitting.  This is reflected in time-domain
phenomena ranging from the suppression of tunneling to a strong increase in its
rate and to complicated quantum beats \cite{Latka1994b}.  Singlet-doublet
crossings, in turn, drastically change the quasienergy scales and replace the
two-level by a three-level structure.

%----------------------------------------------------------------------
\subsubsection{Three-level crossings}
\label{sec:3s}

\begin{figure}[tb]
\hfill
\includegraphics[height=.4\columnwidth]{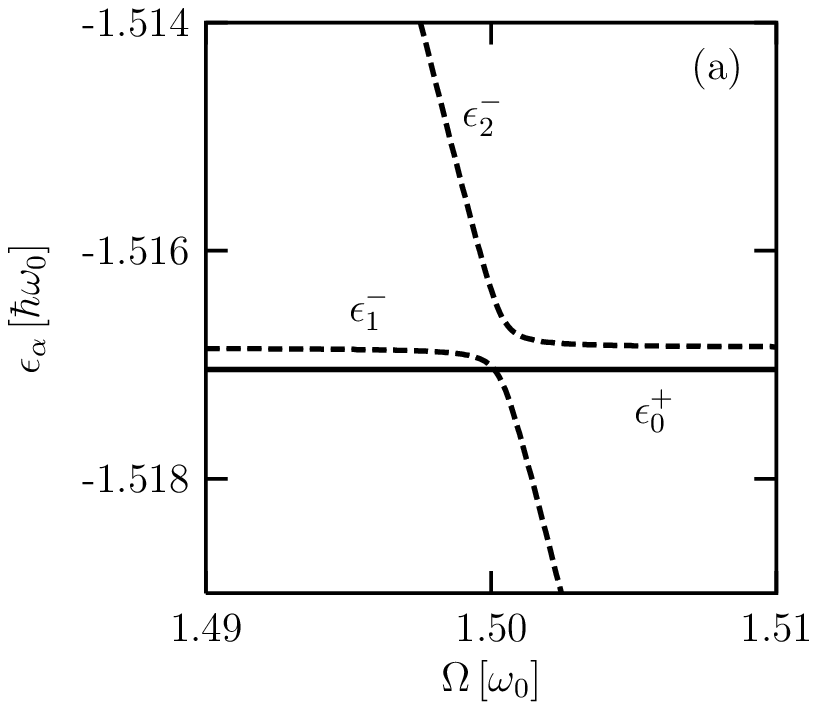}
\hfill
\includegraphics[height=.4\columnwidth]{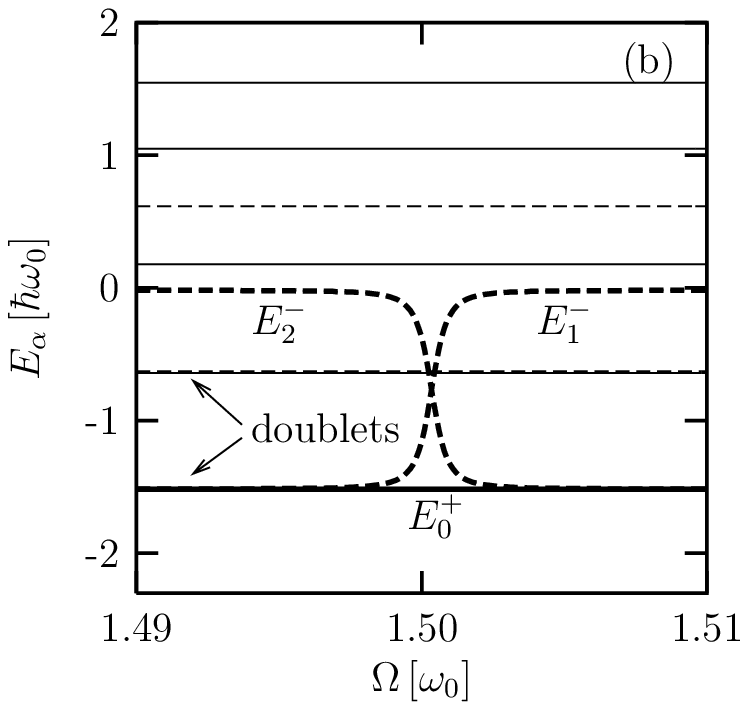}
\hfill
\caption{\label{fig:spectrum}
Quasienergies (a) and mean energies (b) found numerically for the driven double
well potential with $D=E_\mathrm{B}/\hbar\omega_0=2$ and the dimensionless
driving strength $F=10^{-3}$.
Energies of states with even (odd) generalized parity are marked by full
(broken) lines; bold lines (full and broken) correspond to the states
(\ref{eq:psi012}) which are formed from the singlet $|\phi_\mathrm{t}^-\rangle$
and the doublet $|\phi_\mathrm{d}^\pm\rangle$.
A driving frequency $\Omega>1.5\,\omega_0$ corresponds to a detuning
$\delta=E_\mathrm{t}^--E_\mathrm{d}^--\hbar\Omega<0$.
}
\end{figure}

A doublet which is driven close to resonance with a singlet can be adequately
described in a three-state Floquet picture.  For a quantitative account of
such crossings and the associated coherent dynamics, and for later reference in
the context of the incoherent dynamics, we shall now discuss them in terms of a
simple three-state model, which has been discussed in the context of chaotic
tunneling \cite{Bohigas1993a,Kohler1998a}.
In order to illustrate the above three-state model and to demonstrate its
adequacy, we have numerically studied a singlet-doublet crossing that occurs
for the double-well potential, Eq.~(\ref{hamiltonian}), with $D=2$, at a driving
frequency $\Omega \approx 1.5\,\omega_0$ and an amplitude $F=0.001$
(Fig.~\ref{fig:spectrum}).
%----------
%----------

Far outside the crossing, we expect the following situation: There is a doublet
(subscript~d) of Floquet states
\begin{equation}
\begin{split}
|\psi_\mathrm{d}^+(t)\rangle
&= \mathrm{e}^{-\mathrm{i}\epsilon_\mathrm{d}^+
    t/\hbar}|\phi_\mathrm{d}^+(t)\rangle , \\ |\psi_\mathrm{d}^-(t)\rangle
&= \mathrm{e}^{-\mathrm{i}(\epsilon_\mathrm{d}^+ +\Delta) t/\hbar}
    |\phi_\mathrm{d}^-(t)\rangle ,
\end{split}
\end{equation}
with even (superscript $+$) and odd ($-$) generalized parity,
respectively, residing on a pair of quantizing tori in one of the well
regions. We have assumed the quasienergy splitting $\Delta =
\epsilon_\mathrm{d}^- - \epsilon_\mathrm{d}^+$ (as opposed to the unperturbed
splitting) to be positive. The global relative phase is chosen such that
the superpositions
\begin{equation}
\label{eq:rightleft}
|\phi_\mathrm{R,L}(t)\rangle = \frac{1}{\sqrt{2}}\left(|\phi_\mathrm{d}^+(t)\rangle
\pm |\phi_\mathrm{d}^-(t)\rangle\right)
\end{equation}
are localized in the right and the left well, respectively, and
tunnel back and forth with a frequency $\Delta/\hbar$.
 
As the third player, we introduce a Floquet state
\begin{equation}
|\psi_\mathrm{t}^-(t)\rangle
= \mathrm{e}^{-\mathrm{i}(\epsilon_\mathrm{d}^+ + \Delta + \delta)t/\hbar}
|\phi_\mathrm{t}^-(t)\rangle,
\end{equation}
located mainly at the top of the barrier (subscript t), so that its
time-periodic part $|\phi_\mathrm{t}^-(t)\rangle$ contains a large number of
harmonics.  Without loss of generality, its parity is fixed to be odd.
Note that $|\phi_\mathrm{d}^\pm(t)\rangle$ are in general not eigenstates
of the static part of the Hamiltonian, but exhibit for sufficiently strong
driving already a non-trivial, $T$-periodic time-dependence.
For the quasienergy, we assume that $\epsilon_\mathrm{t}^- =
\epsilon_\mathrm{d}^+ + \Delta + \delta = \epsilon_\mathrm{d}^- +
\delta$, where the detuning $\delta=E_\mathrm{t}^--E_\mathrm{d}^--\hbar\Omega$
serves as a measure of the distance from the crossing.  The mean energy of
$|\psi_\mathrm{t}^-(t)\rangle$ lies approximately by $\hbar\Omega$ above the
doublet such that $|E_\mathrm{d}^- - E_\mathrm{d}^+| \ll E_\mathrm{t}^- -
E_\mathrm{d}^\pm$.
 
In order to model an avoided crossing between $|\phi_\mathrm{d}^-\rangle$ and
$|\phi_\mathrm{t}^-\rangle$, we suppose that there is a non-vanishing fixed matrix
element
\begin{equation}
b = \frac{1}{T}\int_0^T\mathrm{d}t\,\langle\phi_\mathrm{d}^-|H_\mathrm{DW}|\phi_\mathrm{t}^-\rangle > 0.
\end{equation}
For the singlet-doublet crossings under study, we typically find
that $\Delta \lesssim b \ll \hbar\Omega$. Neg\-lecting the coupling with all
other states, we model the system by the three-state (subscript 3s) Floquet
Hamiltonian \cite{Kohler1998a,Hanggi1999a}
\begin{equation}
{\cal H}_\mathrm{3s} = \epsilon_\mathrm{d}^+
+\left(\begin{array}{ccc}
        0 & 0      &                     0 \\
        0 & \Delta &                     b \\
        0 & b      & \Delta+\delta
\end{array}\right)
\label{eq:ham3s}
\end{equation}
in the three-dimensional Hilbert space spanned by $\{|\phi_\mathrm{
r}^+(t)\rangle, |\phi_\mathrm{d}^-(t)\rangle, |\phi_\mathrm{t}^-(t)\rangle\}$. Its
Floquet states are
\begin{eqnarray}
\label{eq:psi012}
|\phi_0^+(t)\rangle
&=& %%% \mathrm{e}^{-\mathrm{i}\epsilon_0^+t/\hbar}
|\phi_\mathrm{d}^+(t)\rangle , \nonumber\\
|\phi_1^-(t)\rangle
&=& %%% \mathrm{e}^{-\mathrm{i}\epsilon_1^- t/\hbar}
    \left( |\phi_\mathrm{d}^-(t)\rangle\cos\beta
    - |\phi_\mathrm{t}^-(t)\rangle\sin\beta \right) , \\
|\phi_2^-(t)\rangle
&=& %%% \mathrm{e}^{-\mathrm{i}\epsilon_2^- t/\hbar}
    \left( |\phi_\mathrm{d}^-(t)
    \rangle\sin\beta + |\phi_\mathrm{t}^-(t)\rangle\cos\beta \right) .\nonumber
\end{eqnarray}
with quasienergies
\begin{equation}
\epsilon_0^+ = \epsilon_\mathrm{d}^+,\quad
\epsilon_{1,2}^- = \epsilon_\mathrm{d}^+ + \Delta +
\frac{1}{2}\delta\mp\frac{1}{2} \sqrt{\delta^2+4b^2},
\end{equation}
and mean energies, neglecting contributions of the matrix element $b$,
\begin{eqnarray}
\label{eq:meanen3s}
E_0^+ &=& E_\mathrm{d}^+ , \nonumber\\
E_1^- &=& E_\mathrm{d}^-\cos^2\beta + E_\mathrm{t}^-\sin^2\beta , \\
E_2^- &=& E_\mathrm{d}^-\sin^2\beta + E_\mathrm{t}^-\cos^2\beta . \nonumber
\end{eqnarray}

The angle $\beta$ describes the mixing between the Floquet states
$|\phi_\mathrm{d}^-\rangle$ and $|\phi_\mathrm{t}^-\rangle$ and is an
alternative measure of the distance to the avoided crossing. By diagonalizing
the matrix (\ref{eq:ham3s}), we obtain
\begin{equation}
2\beta = \arctan\left(\frac{2b}{\delta}\right), \quad
0 < \beta < \frac{\pi}{2}.
\end{equation}
For $\beta \to \pi/2$, corresponding to $-\delta\gg b$, we retain the
situation far right of the crossing, as outlined above, with $|\phi_1^- \rangle
\approx-|\phi_\mathrm{t}^-\rangle$, $|\phi_2^-\rangle \approx |\phi_\mathrm{
d}^-\rangle$. To the far left of the crossing, i.e.\ for $\beta \to 0$ or
$\delta\gg b$, the exact eigenstates $|\phi_1^-\rangle$ and
$|\phi_2^-\rangle$ have interchanged their shape \cite{Latka1994b,Kohler1998a}.
Here, we have $|\phi_1^-\rangle \approx |\phi_\mathrm{d}^-\rangle$ and
$|\phi_2^-\rangle \approx |\phi_\mathrm{t}^-\rangle$.  The mean energy is
essentially determined by this shape of the state, so that there is also an
exchange of $E_1^-$ and $E_2^-$ in an exact crossing, cf.\
Eq.~(\ref{eq:meanen3s}), while $E_0^+$ remains unaffected
(Fig.~\ref{fig:spectrum}b).

To study the dynamics of the tunneling process, we focus on the state
\begin{equation}
|\psi(t)\rangle = \frac{1}{\sqrt{2}}\left(
\mathrm{e}^{-\mathrm{i}\epsilon_0^+ t/\hbar}|\phi_0^+(t)\rangle
+\mathrm{e}^{-\mathrm{i}\epsilon_1^- t/\hbar}|\phi_1^-(t)\rangle\cos\beta
+\mathrm{e}^{-\mathrm{i}\epsilon_2^- t/\hbar}|\phi_2^-(t)\rangle\sin\beta
\right).
\label{eq:tunnelstate}
\end{equation}
It is constructed such that at $t = 0$, it corresponds to the decomposition of
$|\phi_\mathrm{R}\rangle$ in the basis (\ref{eq:psi012}) at finite distance
from the crossing. Therefore, it is initially localized
in the right well and follows the time evolution under the Hamiltonian
(\ref{eq:ham3s}). From Eqs.~(\ref{eq:rightleft}), (\ref{eq:psi012}),
we find the probabilities for its evolving into $|\phi_\mathrm{R}\rangle$,
$|\phi_\mathrm{L}\rangle$, or $|\phi_\mathrm{t}\rangle$, respectively, to be
\begin{eqnarray}
P_\mathrm{R,L}(t)
&=& |\langle\phi_\mathrm{R,L}(t)|\psi(t)\rangle|^2 \nonumber\\
&=& \frac{1}{2}\Bigg(1 \pm \left[\cos\frac{(\epsilon_1^--\epsilon_0^+)t}{\hbar}
    \cos^2\beta+
\cos\frac{(\epsilon_2^--\epsilon_0^+)t}{\hbar}\sin^2\beta\right]
    \nonumber\\
&&  {} +\left[\cos\frac{(\epsilon_1^--\epsilon_2^-)t}{\hbar}-1\right]
    \cos^2\beta\sin^2\beta\Bigg), \label{eq:tun3s} \\
P_\mathrm{t}(t) &=& |\langle\phi_\mathrm{t}(t)|\psi(t)\rangle|^2 \nonumber
= \left[1-\cos\frac{(\epsilon_1^--\epsilon_2^-)t}{\hbar}\right]
    \cos^2\beta\sin^2\beta. \nonumber
\end{eqnarray}
At sufficient distance from the crossing, there is only little mixing between
the doublet and the resonant states, i.e., $\sin\beta\ll 1$ or $\cos\beta\ll
1$.  The tunneling process then follows the familiar two-state dynamics
involving only $|\phi_\mathrm{d}^+\rangle$ and $|\phi_\mathrm{d}^-\rangle$,
with tunnel frequency $\Delta/\hbar$. Close to the avoided crossing,
$\cos\beta$ and $\sin\beta$ are of the same order of magnitude, and
$|\phi_1^-\rangle$, $|\phi_2^-\rangle$ become very similar to one another.
Each of them has now support at the barrier top and in the well region, they
are of a hybrid nature. Here, the tunneling involves all the three states and
must be described at least by a three-level system. The exchange of probability
between the two well regions proceeds via a ``stop-over'' at hte top of the
barrier.

%----------------------------------------------------------------------
\section{Dissipative tunneling}

The small energy scales associated with tunneling make it extremely sensitive
to any loss of coherence. As a consequence, the symmetry underlying the
formation of tunnel doublets is generally broken, and an additional energy
scale is introduced, the effective finite width attained by each discrete
level.  As a consequence, the familiar way tunneling fades away in the presence
of dissipation on a time scale $t_\mathrm{coh}$.  In general, this time scale
gets shorter for higher temperatures, reflecting the growth of the transition
rates.  However, there exist counterintuitive effects: in the vicinity of an
exact crossing of the ground-state doublet, coherence can be stabilized with
higher temperatures \cite{Dittrich1993a} until levels outside the doublet start
to play a role.

As a measure for the coherence of a quantum system we employ in this work the
Renyi entropy \cite{Wehrl1991a}
\begin{equation}
\label{renyi}
S_\alpha=\frac{\ln\mathop\mathrm{tr}\rho^\alpha}{1-\alpha} . 
\end{equation}
In our numerical studies we will use $S_2$ which is related to the purity
$\mathop{\mathrm{tr}}(\rho^2)$.  It possesses a convenient physical
interpretation: Suppose that $\rho$ describes an incoherent mixture of $n$
states with equal probability, then $\mathop{\mathrm{tr}}(\rho^2)$ reads $1/n$
and one accordingly finds $S_2=\ln n$.

%---------------------------------------------
\subsubsection{Floquet-Markov master equation}
To achieve a microscopic model of dissipation, we couple the driven bistable system
(\ref{hamiltonian}) bilinearly to a bath of non-interacting harmonic oscillators
\cite{Magalinskii1959a,Caldeira1983a,QTAD}.  The total Hamiltonian of system
and bath is then given by
\begin{equation}
H(t) = H_\mathrm{DW}(t)+
\sum_{\nu=1}^\infty \left(\frac{p_\nu^2}{2m_\nu}+\frac{m_\nu}{2}\omega_\nu^2
\left(x_\nu - \frac{g_\nu}{m_\nu\omega_\nu^2}x \right)^2 \right).
\end{equation}
Due to the bilinearity of the system-bath coupling, one can eliminate
the bath variables to get an exact, closed integro-differential equation for
the reduced density matrix $\rho(t)=\mathrm{tr}_\mathrm{B}\rho_\mathrm{total}(t)$. It
describes the dynamics of the central system, subject to dissipation.

In the case of weak coupling, such that the dynamics is predominatly coherent,
the reduced density operator obeys in good approximation a Markovian master equation.
The Floquet states $|\phi_\alpha(t)\rangle$ form then a well-adapted basis
set for a decomposition that allows for an efficient numerical treatment.  If
the spetral density of the bath influence is ohmic \cite{Caldeira1983a,
QTAD}, the resulting master equation reads \cite{Blumel1989a,Kohler1997a}
\begin{equation}
\dot\rho_{\alpha\beta}(t)
=-\frac{\mathrm{i}}{\hbar}(\epsilon_\alpha-\epsilon_\beta)\rho_{\alpha\beta}(t)
+\sum_{\alpha'\beta'} {\cal L}_{\alpha\beta,\alpha'\beta'}\rho_{\alpha'\beta'}.
\label{mastereq}
\end{equation}
The time-independent dissipative kernel
\begin{eqnarray}
{\cal L}_{\alpha\beta,\alpha'\beta'}
&=& \sum_k \left( N_{\alpha\alpha',k} + N_{\beta\beta',k}\right)
    X_{\alpha\alpha',k} X_{\beta'\beta,-k}
\nonumber\\
&& -\delta_{\beta\beta'}\sum_{\beta'',k}
   N_{\beta''\alpha',k}X_{\alpha\beta'',-k}
   X_{\beta''\alpha',k}
\label{MasterEquationRWA}\\
&& -\delta_{\alpha\alpha'}\sum_{\alpha''k}
   N_{\alpha''\beta',k}
   X_{\beta'\alpha'',-k}X_{\alpha''\beta,k}
\nonumber
\end{eqnarray}
is given by the Fourier coefficients of the position matrix elements,
\begin{equation}
X_{\alpha\beta,k}
= \frac{1}{T}\int_0^T\mathrm{d}t\,\mathrm{e}^{-\mathrm{i}k\Omega t}
\langle\phi_\alpha(t)|x|\phi_\beta(t)\rangle\rangle = X_{\beta\alpha,-k}^*
\end{equation}
and the coefficients
\begin{equation}
N_{\alpha\beta,k} = N(\epsilon_\alpha-\epsilon_\beta+k\hbar\Omega),\quad
N(\epsilon) = \frac{m\gamma\epsilon}{\hbar^2}
\frac{1}{\mathrm{e}^{\epsilon/k_\mathrm{B}T}-1}
\end{equation}
which consist basically of the spectral density times the thermal occupation of
the bath.

%-----------------------------------------
\subsubsection{Dissipative time evolution}
We have studied dissipative tunneling at the particular singlet-doublet
crossing introduced in Sec.~\ref{sec:3s} (see Fig.~\ref{fig:spectrum}).  The
time evolution has been computed numerically by integrating the master equation
(\ref{mastereq}).  As initial condition, we have chosen the density operator
$\rho(0)=|\phi_\mathrm{L}\rangle\langle\phi_\mathrm{L}|$, i.e.\ a pure state
located in the left well.
 
%----------
\begin{figure}[tb]
\hfill
\includegraphics[height=.4\columnwidth]{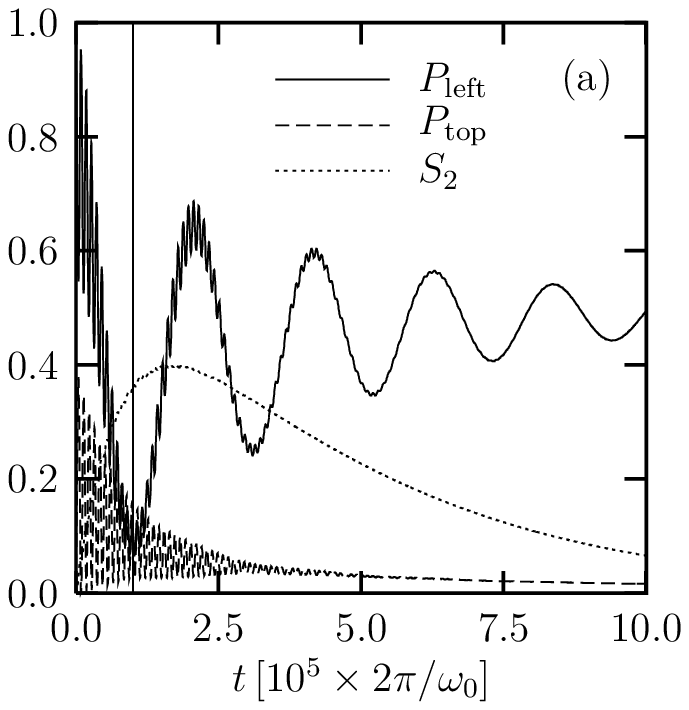}
\hfill
\includegraphics[height=.4\columnwidth]{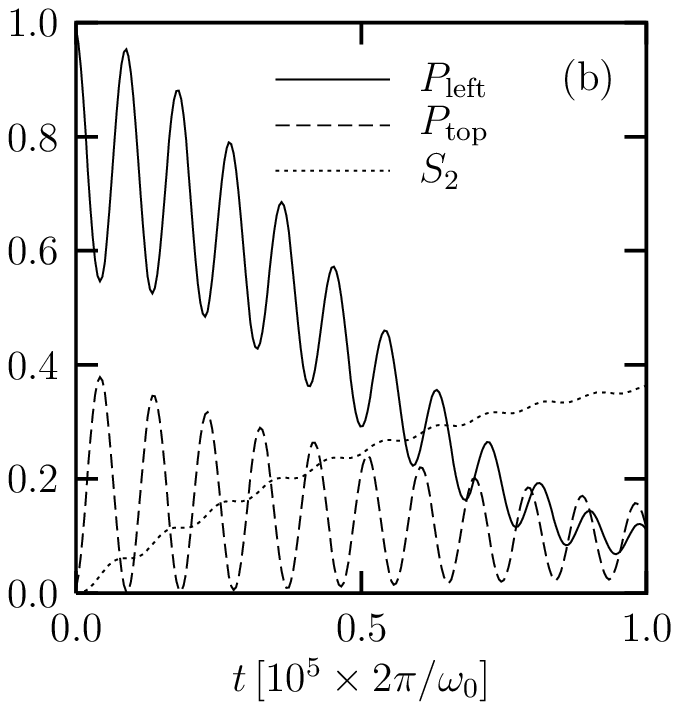}
\hfill
\caption{\label{fig:dynamics}
Time evolution of the state $|\phi_\mathrm{L}\rangle$ at the center of the
singlet-doublet crossing found for $D=2$, $F=10^{-3}$, and
$\Omega=1.5\,\omega_0$.
The full line depicts the return probability and the broken line the occupation
probability of the state at the top of the barrier.  The dotted line marks
the Renyi entropy $S_2$.
Panel (b) is a blow-up of the marked region on the left of panel (a).}
\end{figure}%
%----------
In the vicinity of a singlet-doublet crossing, the tunnel splitting
increases and during the tunneling, the singlet $|\phi_\mathrm{t}\rangle$
at the top of the barrier becomes populated
periodically with frequency $|\epsilon_2^--\epsilon_1^-|/\hbar$, cf.\
Eq.~(\ref{eq:tun3s}) and Fig.~\ref{fig:dynamics}b. The large mean energy of this
singlet results in an enhanced entropy production at times when it is well
populated (dashed and dotted line in Fig.~\ref{fig:dynamics}b). For the
relaxation towards the asymptotic state, also the slower transitions within
doublets are relevant.
Therefore, the corresponding time scale can be much larger than
$t_\mathrm{coh}$ (dotted line in Fig.~\ref{fig:dynamics}a).

\begin{figure}[tb]
\hfill
\includegraphics[height=.4\columnwidth]{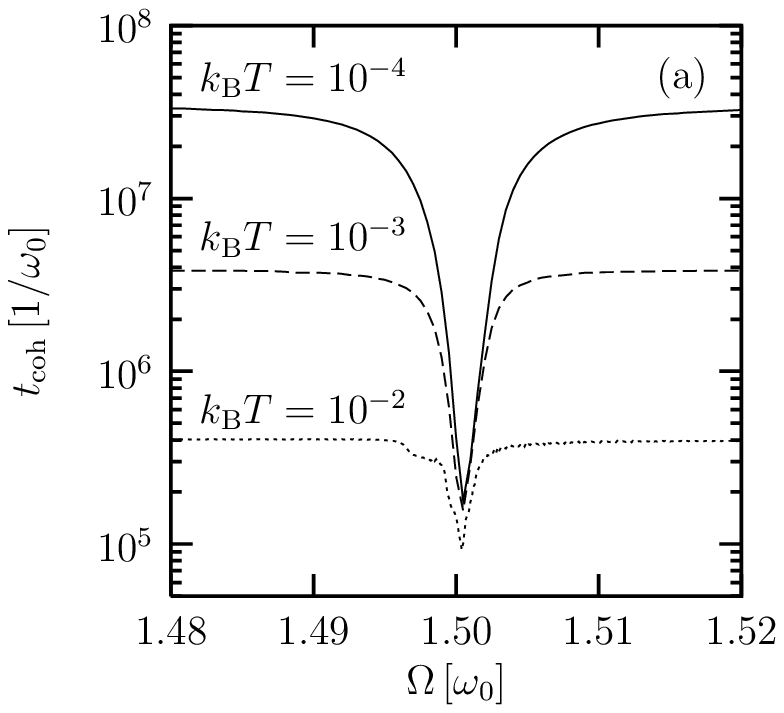}
\hfill
\includegraphics[height=.4\columnwidth]{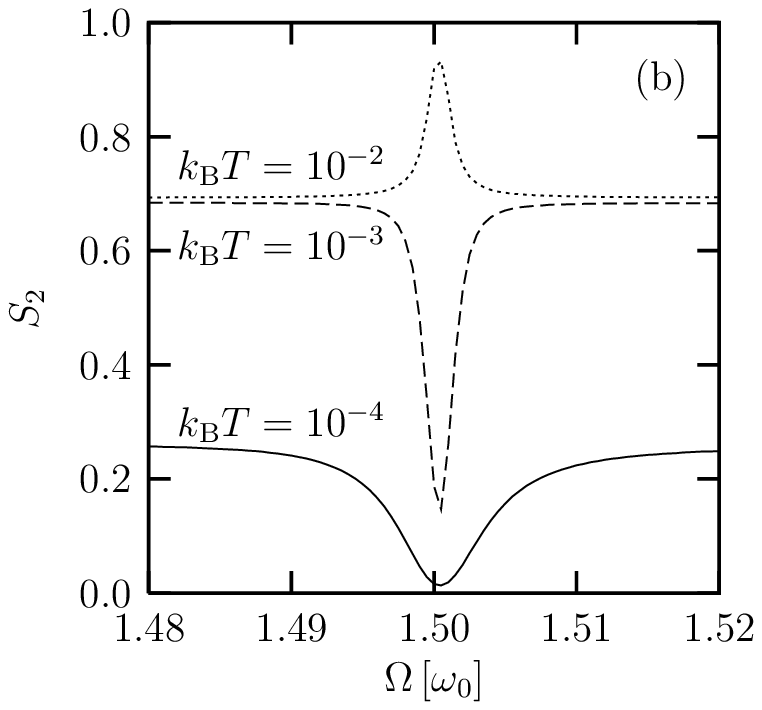}
\hfill
\caption{\label{fig:decoherence}
Decoherence time (a) and Renyi entropy $S_2$ of the asymptotic state (b) in the
vicinity of the singlet-doublet crossing for $D=2$, $F=10^{-3}$, and
$\Omega=1.5\,\omega_0$.
The temperature is given in units of $\hbar\omega_0$.
}
\end{figure}

To obtain quantitative estimates for the dissipative time scales, we
approximate $t_\mathrm{coh}$ by the growth of the Renyi entropy,
averaged over a time $t_\mathrm{p}$,
\begin{equation}
\frac{1}{t_\mathrm{coh}}
= \frac{1}{t_\mathrm{p}}\int_0^{t_\mathrm{p}} \mathrm{d}t' \dot S_2(t')
= \left.\frac{1}{t_\mathrm{p}}\Big( S_2(t_p)-S_2(0))\Big)
\right. .
\end{equation}
Because of the stepwise growth of the Renyi entropy (Fig.~\ref{fig:dynamics}b),
we have chosen the propagation time $t_\mathrm{p}$ as an $n$-fold multiple of
the duration $2\pi\hbar/|\epsilon_2^- - \epsilon_1^-|$ of a tunnel cycle.  For
this procedure to be meaningful, $n$ should be so large that the Renyi entropy
increases substantially during the time $t_\mathrm{p}$ (in our numerical
studies from zero to a value of approximately 0.2).  We find that at the center
of the avoided crossing, the decay of coherence, respectively the entropy
growth, becomes much faster and is
essentially independent of temperature (Fig.~\ref{fig:decoherence}a).  At a
temperature $k_\mathrm{B}T=10^{-4}\hbar\omega_0$ it is enhanced by three orders
of magnitude.  This indicates that transitions from states with mean energy far
above the ground state play a crucial role.
%----------
%----------

As the dynamics described by the master equation (\ref{mastereq}) is
dissipative, it converges in the long-time limit to an asymptotic state
$\rho_\infty(t)$. In general, this attractor remains time dependent but shares
the symmetries of the central system, i.e.\ here, periodicity and generalized
parity. However, the coefficients (\ref{MasterEquationRWA}) of the master
equation for the matrix elements $\rho_{\alpha\beta}$ are time independent and
so the asymptotic solution also is.  The explicit time dependence of the
attractor has been effectively eliminated by the use of a Floquet basis.
 
To gain some qualitative insight into the asymptotic solution, we focus on
the diagonal elements
\begin{equation}
{\cal L}_{\alpha\alpha,\alpha'\alpha'}
=2\sum_n N_{\alpha\alpha',n}|X_{\alpha\alpha',n}|^2,\quad \alpha\neq\alpha',
\label{fullRWA}
\end{equation}
of the dissipative kernel. They give the rates of direct transitions from
$|\phi_{\alpha'}\rangle$ to $|\phi_\alpha\rangle$. Within a golden rule
description, these were the only non-vanishing contributions to the master
equation to affect the diagonal elements $\rho_{\alpha\alpha}$ of the density
matrix.
 
In the case of zero driving amplitude, the Floquet states $|\phi_\alpha\rangle$
reduce to the eigenstates of the undriven Hamiltonian $H_\mathrm{DW}$. The only
non-vanishing Fourier component is then $|\phi_{\alpha,0}\rangle$, and the
quasienergies $\epsilon_\alpha$ reduce to the corresponding eigenenergies
$E_\alpha$.  Thus ${\cal L}_{\alpha\alpha,\alpha'\alpha'}$ only consists of a
single term proportional to $N(E_\alpha-E_{\alpha'})$.  It
describes two kinds of thermal transitions: decay to states with lower energy
and, if the energy difference is less than $k_\mathrm{B}T$, thermal activation
to states with higher energy. The ratio of the direct transitions forth and
back then reads
\begin{equation}
\frac{{\cal L}_{\alpha\alpha,\alpha'\alpha'}}{{\cal L}_{\alpha'\alpha'
,\alpha\alpha}}
=\exp\left(-\frac{E_\alpha-E_{\alpha'}}{k_\mathrm{B}T}\right).
\end{equation}
We have detailed balance and therefore the steady-state solution is
\begin{equation}
\rho_{\alpha\alpha'}(\infty) \sim \mathrm{e}^{-E_\alpha/k_\mathrm{B}T}\,
\delta_{\alpha\alpha'}.
\end{equation}
In particular, the occupation probability decays monotonically with the energy
of the eigenstates. In the limit $k_\mathrm{B}T\to0$, the system tends to occupy
mainly the ground state.
 
For a strong driving, each Floquet state $|\phi_\alpha\rangle$ contains a large
number of Fourier components and ${\cal L}_{\alpha\alpha,\alpha'\alpha'}$ is
given by a sum over contributions with quasienergies $\epsilon_\alpha -
\epsilon_{\alpha'} + k\hbar\Omega$. Thus, a decay to states with ``higher''
quasienergy (recall that quasienergies do not allow for a global ordering)
becomes possible due to terms with $k<0$. Physically, it amounts to an
incoherent transition under absorption of driving-field quanta.
Correspondingly, the system tends to occupy Floquet states comprising many
Fourier components with low index $k$. According to
Eq.~(\ref{eq:meanen:fourier}), these states have a low mean energy.
 
The effects under study are found for a driving with a frequency of the order
$\omega_0$. Thus, for a quasienergy doublet, not close to a crossing, we have
$|\epsilon_\alpha-\epsilon_{\alpha'}| \ll \hbar\Omega$, and ${\cal
L}_{\alpha'\alpha',\alpha\alpha}$ is dominated by contributions with $n<0$,
where the splitting has no significant influence.  However, except for the
tunnel splitting, the two partners in the quasienergy doublet are almost
identical.  Therefore, with respect to dissipation, both should behave
similarly. In particular, one expects an equal population of the doublets even
in the limit of zero temperature in contrast to the time-independent case.

In the vicinity of a singlet-doublet crossing the situation is more subtle.
Here, the odd partner, say, of the doublet mixes with the singlet, cf.\
Eq.~(\ref{eq:psi012}), and thus acquires components with higher energy.  Due to
the high mean energy $E_\mathrm{t}^-$ of the singlet, close to the top of the
barrier, the decay back to the ground state can also proceed indirectly via
other states with mean energy below $E_\mathrm{t}^-$.  Thus, $|\phi_1^-\rangle$
and $|\phi_2^-\rangle$ are depleted and mainly $|\phi_0^+\rangle$ will be
populated.  However, if the temperature is significantly above the splitting
$2b$ at the avoided crossing, thermal activation from $|\phi_0^+\rangle$ to
$|\phi_{1,2}^-\rangle$, accompanied by depletion via the states below
$E_\mathrm{ t}^-$, becomes possible.  Asymptotically, all these states
become populated in a cyclic flow.

In order to characterize the coherence of the asymptotic state, we use again the
Renyi entropy (\ref{renyi}).  According to the above scenario, we expect $S_2$
to assume the value $\ln2$, in a regime with strong driving but preserved
doublet structure, reflecting the incoherent population of the ground-state
doublet. In the vicinity of the singlet-doublet crossing where the doublet
structure is dissolved, its value should be of the order unity for temperatures
$k_\mathrm{B}T\ll 2b$ and much less than unity for $k_\mathrm{B}T\gg 2b$
(Fig.~\ref{fig:decoherence}b). This means that the crossing of the singlet with
the doublet leads asymptotically to an improvement of coherence if the
temperature is below the splitting of the avoided crossing.  For temperatures
above the splitting, the coherence becomes derogated. This phenomenon compares
to chaos-induced coherence or incoherence, respectively, found in
Ref.~\cite{Kohler1998a} for dissipative chaos-assisted tunneling.

%----------------------------------------------------------------------
\section{Conclusions}

For the generic situation of the dissipative quantum dynamics of a particle in
a driven double-well potential, resonances play a significant role for the
loss of coherence.  The influence of states with higher energy alters the
splittings of the doublets and thus the tunneling rates.  We have studied
decoherence in the vicinity of crossings of singlets with tunnel doublets under
the influence of an environment.  As a simple intuitive model to compare
against, we have constructed a three-state system which in the case of
vanishing dissipation, provides a faithful description of an isolated
singlet-doublet crossing.
The center of the crossing is characterized by a strong mixing
of the singlet with one state of the tunnel doublet.  The high mean energy of
the singlet introduces additional decay channels to states outside the
three-state system. Thus, decoherence becomes far more effective and,
accordingly, coherent oscillations fade away on a much shorter time scale.

%----------------------------------------------------------------------
\renewcommand\bibname{References}


\begin{thebibliography}[1]{99}\label{sect:labib}

\bibitem{Dittrich1993a} T.~Dittrich, B.~Oelschl\"agel, and P.~H\"anggi,
Europhys.~Lett.\ {\bf 22},  5 (1993).

\bibitem{Grifoni1998a} M.~Grifoni and P.~H\"anggi, Phys.~Rep.\ {\bf 304},  229
(1998).

\bibitem{Kohler1998a} S.~Kohler, R.~Utermann, P.~H\"anggi, and T.~Dittrich,
Phys.~Rev.~E {\bf 58}, 7219  (1998).

\bibitem{Hanggi1999a} P. H\"anggi, S. Kohler, and T. Dittrich,
\textit{Driven Tunneling: Chaos and Decoherence},
in Statistical and Dynamical Aspects of Mesoscopic Systems, Lecture Notes in
Physics 547, p.125-157 (Springer, 2000)

\bibitem{Shirley1965a} J.~H.~Shirley, Phys.~Rev.\ {\bf 138},  B979  (1965).

\bibitem{Sambe1973a} H.~Sambe, Phys.~Rev.~A {\bf 7},  2203  (1973).
 
\bibitem{Manakov1986a} N.~L.~Manakov, V.~D.~Ovsiannikov, and L.~P.~Rapoport,
Phys.~Rep.\ {\bf 141}, 319  (1986).
 
\bibitem{QTAD} T.~Dittrich, P.~ H\"anggi, G.-L.~ Ingold, B.~ Kramer,
G.~ Sch\"on, and W.~ Zwerger, {\em Quantum Transport and Dissipation}
(Wiley-VCH, Weinheim, 1998).

\bibitem{Grossmann1991a} F.~Grossmann, T.~Dittrich, P.~Jung, and
P.~H\"anggi, Phys.~Rev.~Lett.\ {\bf 67}, 516  (1991).

\bibitem{Grossmann1991b} F.~Grossmann, P.~Jung, T.~Dittrich, and
P.~H\"anggi, Z. Phys. B, {\bf 84}, 315 (1991).

\bibitem{Grossmann1992a} F.~Grossmann and P.~H\"anggi, Europhys.~Lett.\ {\bf
18},  571  (1992).

\bibitem{Latka1994b}
M.~Latka, P.~Grigolini, and B.~J.~West, Phys.~Rev.~A {\bf 50},  1071  (1994).

\bibitem{Bohigas1993a}
O.~Bohigas, S.~Tomsovic, and D.~Ullmo, Phys.~Rep.\ {\bf 223},  43  (1993).

\bibitem{Wehrl1991a} A. Wehrl, Rep. Math. Phys {\bf 30}, 119 (1991).

\bibitem{Magalinskii1959a} V.~B.~Magalinski{\chardef\i="10 \u\i},
Zh.~Eksp.~Teor.~Fiz.\ {\bf 36},  1942 (1959), [Sov.~Phys.~JETP {\bf 9}, 1381
(1959)].

\bibitem{Caldeira1983a} A.~O.~Caldeira and A.~L.~Leggett, Ann.~Phys.~(N.Y.)
{\bf 149},  374  (1983); erratum: Ann. Phys. (N.Y.) {\bf 153}, 445 (1984).

\bibitem{Blumel1989a}
R.~Bl\"umel {\it et~al.}, Phys.~Rev.~Lett.\ {\bf 62},  341  (1989).

\bibitem{Kohler1997a} S.~Kohler, T.~Dittrich, and P.~H\"anggi, Phys.~Rev.~E
{\bf 55},  300  (1997).

\end{thebibliography}
\end{document}